\documentclass[12pt]{spieman} 

\usepackage{amsmath,amsfonts,amssymb}
\usepackage{graphicx}
\usepackage{setspace}
\usepackage{tocloft}
\usepackage{lineno}
\usepackage{subcaption}
\usepackage{listings}
\title{Analysis of Angular-Differential Post-Processing Algorithms for Exoplanet Direct Detection with a Photonic Lantern Nuller}

\author[a,b,c,*]{Suvinay Goyal} 
\author[a,d]{Yinzi Xin} 
\author[a]{Nemanja Jovanovic} 
\author[a]{Dimitri Mawet} 
\author[e]{Michael P. Fitzgerald} 
\affil[a]{Department of Astronomy, California Institute of Technology, 1200 E California Blvd, Pasadena, CA, 91125, USA}
\affil[b]{Department of Astronomy, University of Illinois Urbana-Champaign, Urbana, IL, 61801, USA}
\affil[c]{Department of Physics, California Institute of Technology, 1200 E California Blvd, Pasadena, CA, 91125, USA}
\affil[d]{Leiden Observatory, Niels Bohrweg 2, 2333 CA Leiden, The Netherlands}
\affil[e]{Department of Physics \& Astronomy, 430 Portola Plaza, University of California, Los Angeles, CA 90095, USA}

\cftpagenumbersoff{figure}
\cftpagenumbersoff{table} 
\begin{document} 
\maketitle

\begin{abstract}
Exoplanet research is essential for understanding planetary formation and the potential for life beyond our solar system. The direct imaging method captures exoplanet light while minimizing light from the host star. This is conventionally achieved with a coronagraph, which allows detailed characterization of planetary atmospheres and features. The Photonic Lantern Nuller (PLN) is an innovative instrument designed for the direct detection of closely orbiting exoplanets within the inner working angle of standard coronagraphs. Unlike traditional coronagraphs, where the planet's signal is usually rotationally invariant, with the same point-spread-function at different position angles, and which also overlaps minimally with the residual stellar signal, data from a PLN consist of a one-dimensional collection of points that do not have rotational invariance and overlap significantly with the residual starlight arising from wavefront errors. Exploiting angular diversity to subtract these stellar residuals with the PLN thus requires adapting the Angular Differential Imaging (ADI) technique for use with non-rotationally invariant planet signals at close separations, where strong self-subtraction effects occur. We reformulate ADI using principal component analysis to develop a method to extract spatial parameters of exoplanets from simulated one-dimensional PLN data. We test two variations of ADI on simulated data and show that injecting an antiplanet signal before stellar estimation helps localize the planet due to self-subtraction at lower separations.
\end{abstract}


{\noindent \footnotesize\textbf{*}Suvinay Goyal,  \linkable{suvinay2@illinois.edu} }

\begin{spacing}{2}   

\section{Introduction}
To date, high-contrast imaging has been most successful in observing self-luminous, massive planets at distances $>$ 10 AU from their stellar hosts \cite{NASA_Exoplanets}. The detection and characterization of these planets is enabled by coronagraphs, which block starlight from reaching the final science detector, dramatically reducing the photon noise from the star. Coronagraph designs are widely varied and include the Lyot coronagraph \cite{Lyot_1939}, vortex coronagraph \cite{Palacios_2005, Mawet_2005, Mawet_2013}, shaped pupil coronagraph \cite{Kasdin:05, Carlotti_2011}, achromatic interferometric coronagraph \cite{Baudoz_2000}, and more \cite{Soummer_2003a, Soummer_2003b, Rouan_2000, Vanderbei_2003}. Coronagraphs typically operate at working angles of greater than a few $\lambda/D$, a characteristic spatial scale set by diffraction, defined in terms of the operating wavelength ($\lambda$) and the telescope diameter ($D$).

\begin{figure}[H]
    \centering
    \begin{subfigure}[t]{0.49\textwidth}
        \centering
        \includegraphics[width=\linewidth,
                     trim=0 0 0 0,clip]{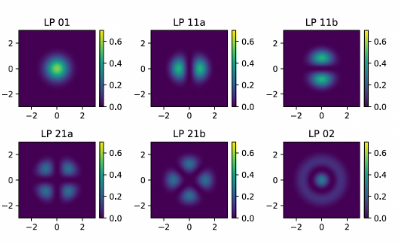}
        \caption{}
        \label{fig:PLNNulling}
    \end{subfigure}
    \begin{subfigure}[t]{0.49\textwidth}
        \centering
        \includegraphics[width=\linewidth]{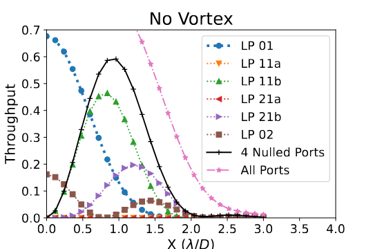}
        \caption{}
        \label{fig:PLNThroughput}
    \end{subfigure}
    \caption{PLN Overview: (a) PLN Selective Nulling: Throughput maps for each port of an ideal six-port mode-selective photonic lantern, spanning from -3 $\lambda/D$ to 3 $\lambda/D$ in each direction.(b) PLN Throughput: Throughput of the Photonic Lantern Nuller. The PLN used throughout this work is depicted by the black curve, which has 4 out of the 6 ports nulled (all except the first and last port), and the throughput is shown peaking around 1 $\lambda/D$, much smaller than the average coronagraph throughput peaks. Figure adapted from Xin et al. \cite{Xin_2022}}
    \label{fig:PLNInfo}
\end{figure}

The Photonic Lantern Nuller (PLN) was designed as an instrument concept \cite{Xin_2022} to help observe closer and fainter off-axis exoplanets to their stellar hosts, at separations at and within $1\lambda/D$ that coronagraphs cannot effectively reach. The PLN uses a mode-selective photonic lantern (MSPL) \cite{LeonSaval_MSPL} to capture a small field of view around the star. For a PLN implemented with a 6-port MSPL, four of the ports have anti-symmetric modes, which, when combined with the symmetric stellar electric field, causes the starlight to be cancelled through destructive interference. This process nulls the starlight, reducing its contribution to photon noise and allowing for the detection of fainter exoplanets at close separations (Fig.~\ref{fig:PLNInfo}). A particularly useful application of the PLN is to characterize objects detected using conventional coronagraphs, but at longer wavelengths. Planets that are observable just outside the inner working angle of a coronagraph in visible light would fall within the inner working angle at longer wavelengths. In this regime, one could hope to detect different molecular features that help provide more robust atmospheric constraints.

Post-processing the high-contrast imaging data is an integral part of exoplanet detection and characterization. It contributes to high-order speckle attenuation for the imaged exoplanets by removing the stellar speckles that arise from optical aberrations, resulting in improved contrast. Post-processing techniques developed for conventional coronagraphs include reference differential imaging (RDI) \cite{Smith_and_Terrile_1984}, in which data of a point-source reference star is used to estimate the stellar speckles, and angular differential imaging (ADI) \cite{marois_adi}, which exploits the rotation of the sky relative to the instrument to estimate the stellar speckles. Although RDI can be directly applied to the PLN by observing a reference star, it incurs additional observational overheads and can suffer from the effects of telescope instability, differences in star brightness and color between the reference and target star, and (in the case of spectroscopic data), uncorrected or imperfectly corrected differential atmospheric dispersion. Meanwhile, conventional ADI relies on the rotational invariance of the planet signal as well as the ability to rotate a 2D coronagraphic image, and thus is not compatible with PLN data without alterations.

When observing with ground-based telescopes, it is, in principle, also possible to observe in a mode that tracks the field instead of the telescope pupil, thus fixing the location of the planet. However, on many of the 6-10m class telescopes, the field-tracking mode of observation causes the telescope pupil to rotate relative to the downstream instruments (including the adaptive optics system) over time. This is very undesirable if the instrument has been designed for a fixed pupil shape, such is the case with the Subaru Coronagraphic Extreme Adaptive Optics (SCExAO) instrument \cite{Jovanovic_2015}, the only on-sky instrument that currently includes a PLN. The rotation of the pupil typically also increases the dynamic rate of optical aberrations seen by the instrument, making it much more difficult to calibrate the wavefront or to apply dark-hole techniques that improve the contrast \cite{Giveon_2007, Haffert_2023}. Furthermore, this observing mode would not provide any angular diversity, resulting in poorer constraints on the stellar signal as well as increased degeneracy in the planet's location \cite{Xin_2022}. Therefore, an ADI-inspired algorithm is needed to process ground-based PLN data taken in pupil-tracking mode and leverage the angular diversity that is naturally obtained as a result. This algorithm would also be applicable to a PLN on a space telescope, where telescope rolls can be used to provide angular diversity.

In this work, we reformulate the ADI algorithm for use with the PLN. We perform a statistical comparison of several variations of the algorithm as applied to simulated data to try to determine the most sensitive and statistically robust method.

\section{Background}
\label{sec:Background}
\subsection{Photonic Lantern Nuller}

\begin{figure}[H]
    \centering
    \begin{subfigure}[t]{0.45\textwidth}
        \centering
        \includegraphics[width=\linewidth]{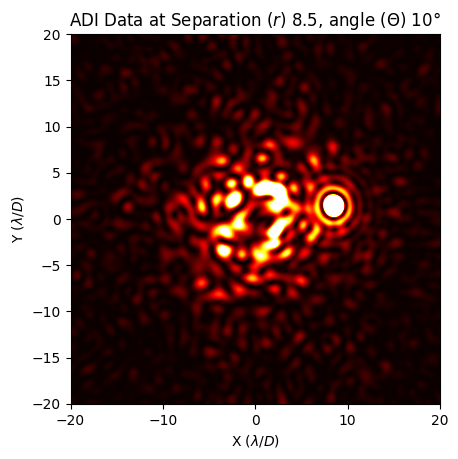}
        \label{fig:CoroPSF}
        \caption{}
    \end{subfigure}
    \begin{subfigure}[t]{0.45\textwidth}
        \centering
        \includegraphics[width=\linewidth]{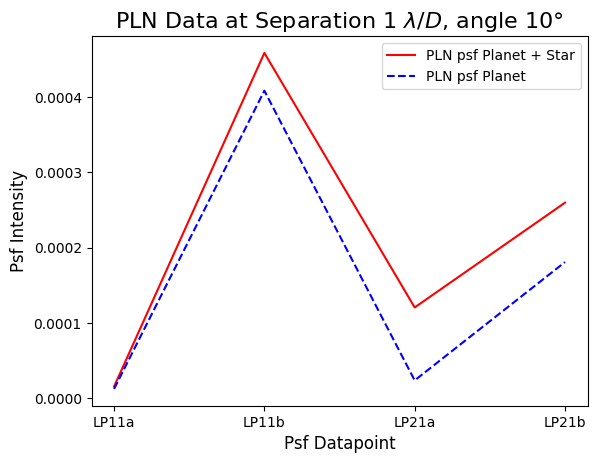}
        \caption{}
        \label{fig:PLN_PSF}
    \end{subfigure}
    \caption{Comparison of instrument data: (a) Simulated 2-Dimensional data from a Charge 6 Vortex Coronagraph containing a bright planet PSF against a field of speckles (the intensity contributed by starlight due to wavefront error), (b) Simulated 1-Dimensional data from the four nulled ports of a PLN. The blue dashed line corresponds to the intensity contributed by a bright planet, while the red solid line corresponds to the combined intensity of the same planet with the intensity contributed by starlight leaked due to wavefront error.}
    \label{fig:PSF_comparison}
\end{figure}

A photonic lantern decomposes the incident signal onto the modes of a few-mode fiber (FMF) face and maps each onto individual single-mode fiber (SMF) ports. The spatial variation of the coupling into each port allow us to put limits on the planet’s on sky position and emission within a small field of view around the star. The PLN also cancels out the central starlight through the destructive interference within the anti-symmetric ports of a mode-selective photonic lantern, thus ``nulling" the starlight, allowing us to see fainter exoplanets at closer separations. For a PLN that uses a 6-port mode-selective photonic lantern, four of the ports are nulled: LP11a, LP11b, LP21a, LP21b (Fig.~\ref{fig:PLNInfo}a), so the useful output signal is a collection of 4 data points of light intensity from the 4 nulled ports of the PLN. The throughput curve for the 4 nulled ports of the PLN is shown in Fig.~\ref{fig:PLNInfo}b, overplotted with the individual port throughputs and the combined throughput of all six ports (including the two non-nulled ports). The 4-nulled-port throughput peaks around $0.75$--$1.25\,\lambda/D$, demonstrating the instrument’s sensitivity to planets at small inner working angles. This results in an overall 1D output signal which includes the intensity contributed by the planet and the star, in contrast to the 2D output image from a coronograph (see Fig.~\ref{fig:PSF_comparison}). The simulations shown in Fig. \ref{fig:PSF_comparison} and throughout this work were conducted using the \lstinline{hcipy} package with a circular aperture, with pupil diameter, wavelength, and focal lengths all set to 1. The vortex coronagraphs were simulated using \lstinline{VortexCoronagraph} with the desired charge, along with a circular Lyot Stop of diameter 0.95. The photonic lantern was simulated by calculating the overlap integral of the focal-plane electric field with the linearly polarized (LP) modes, a basis for modes propagating through circular step-index fibers \cite{lp_mode_def}, which also describes the modes of each port of a MSPL. The LP modes are generated using \lstinline{make_LP_modes} with a $V$ number of $1.5\pi$ and a core radius of $1.4 \lambda/D$ (the value that maximizes the planet throughput through the instrument as described in \cite{Xin_2022}). 

The traditional ADI technique relies on two main assumptions about the initial 2-D signals from coronagraphs:

\begin{itemize}
\item The planet's PSF through the coronagraph is rotationally invariant.
\item The planet's PSF does not significantly overlap with the stellar host's PSF at high spatial separations.
\end{itemize}

Based on these assumptions, images are taken by commanding the telescope rotator to follow the pupil rather than the field. In this observing mode, the quasistatic speckle structure remains constant while the off-axis companion rotates through the field of view. Post processing is then done by subtracting the stellar speckles using methods such as median subtraction \cite{marois_adi}, LOCI \cite{Lafrenière_2007}, or KLIP \cite{Soummer_2012}, followed by derotating the reference images to stack up the planet signal, thereby improving the SNR. Within this framework, the spatial structure of the planet PSF is assumed to be effectively preserved across frames, and derotation can be handled as a coordinate transformation in image space.

In contrast, the PLN produces a low-dimensional vector of modal intensities, for which no equivalent geometric derotation exists. The variation of the PLN signal with rotation is governed by the interferometric response of the instrument, and its evolution with rotation is not describable as simple transformation applied to the basis over which the signal is expressed. As a result, the standard image-based derotation paradigm does not directly extend to PLN data. This distinction motivates the need for algorithms that fit the signal as observed in measurement space, rather than relying on geometric alignment.

\subsection{Comparative Problems} 

The overlap between PLN PSFs, compared with that of coronagraphic PSFs at their optimal throughputs, can be characterized by taking normalized inner products of the PSFs from both instruments at different sky angles. This captures the similarity between PSFs at different angular orientations. For the coronagraph, a single PSF at a given separation is rotated to angles $\phi_1$ and $\phi_2$. These rotated PSFs are then flattened and normalized before taking the dot product. For the PLN, the 1D PSFs are inherently functions of their separations and angles. They are normalized and dotted to obtain the correlation.

\begin{figure}[H]
    \centering
    \begin{subfigure}[t]{0.32\textwidth}
        \centering
        \includegraphics[width=\linewidth]{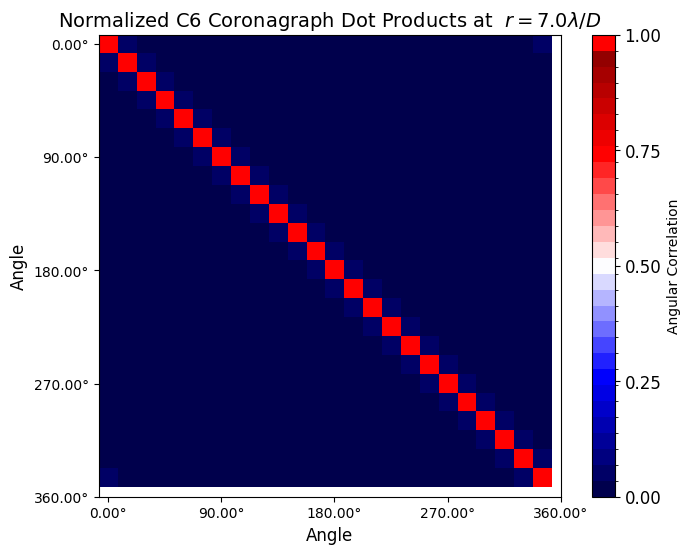}
        \label{fig:CoroIP_7}
        \caption{}
    \end{subfigure}
    \begin{subfigure}[t]{0.32\textwidth}
        \centering
        \includegraphics[width=\linewidth]{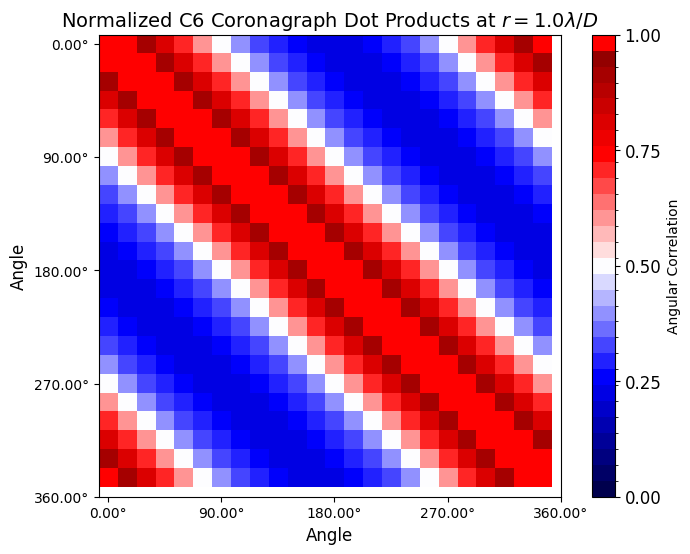}
        \label{fig:CoroIP_1}
        \caption{}
    \end{subfigure}
    \begin{subfigure}[t]{0.32\textwidth}
        \centering
        \includegraphics[width=\linewidth]{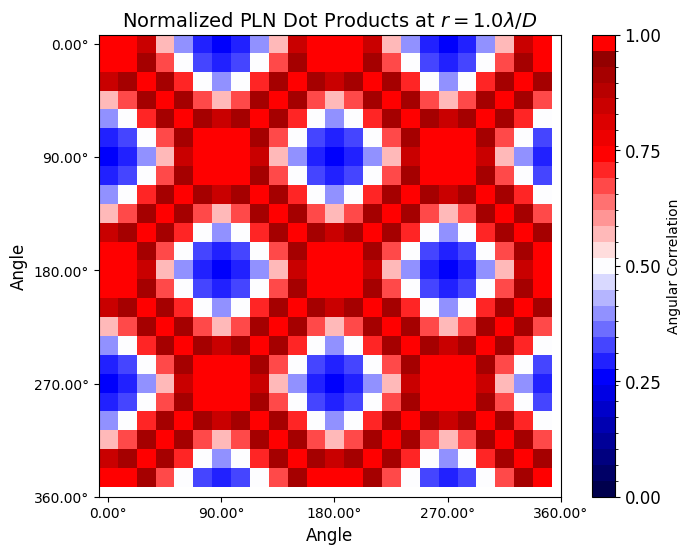}
        \label{fig:PLNIP_1}
        \caption{}
    \end{subfigure}
    \caption{Comparison of planet PSF correlations across rotation angles : (a) The far off-axis (a separation of $7 \lambda/D$) PSFs of a charge 6 vortex coronagraph are generated, ranging in position angle from 0 to $2\pi$. The 2D images are flattened into a 1D array, normalized, and dotted with each other, showing almost no inter-correlations. (b) The close off-axis (a separation of $1 \lambda/D$) PSFs of a charge 6 vortex coronagraph are generated, ranging in position angle from 0 to $2\pi$ (included for comparison with the PLN, even though this is well within the inner working angle of this coronagraph). The 2D images are flattened, normalized, and dotted with each other to show correlations. (c) Normalized dot products between PSFs generated by the mode-selective PLN at its peak throughput separation of $1,\lambda/D$, for position angles ranging from 0 to $2\pi$. Each PSF’s 1D intensity profile is normalized before computing pairwise dot products, revealing a characteristic angular correlation structure. High correlations occur between PSFs at nearby position angles, with correlation decreasing gradually up to an angular separation of $\pi$. The repeating pattern observed at $\phi_{2} + \pi$ for each pair $(\phi_{1}, \phi_{2})$ indicates the intrinsic $180^{\circ}$ spatial degeneracy of the planet signal captured by the PLN.}
    \label{fig:Psf_dot_products}
\end{figure}

For coronagraphs, the correlation between PSFs at nearby angles is low at higher separations but higher at smaller separations. For the PLN, which operates only at small separations, the correlation at nearby positions is still high, but it decreases more rapidly as a function of angle than for a coronagraph operating at the same separation (Fig. ~\ref{fig:Psf_dot_products}).

Since a PLN is designed to operate at smaller separations ($\sim 1 \lambda/D$), it is also necessary to characterize the self-subtraction effect that an ADI-like technique would impart on the planet signal and place it in context to the level of self-subtraction observed in processing coronagraphic data. We can assess the scale of self-subtraction at relevant separations for a PLN relative to charge 2, charge 4, and charge 6 vortex coronagraphs by modeling the subtraction of the planet signal with itself (resulting from median subtraction in this case) and calculating the RMS signal intensity retained in the planet window before and after the subtraction. The simulated planet signal used here is described in Section 4.1, with on-sky coverage, $\Delta \theta = 25^{\circ}$ across 20 frames. For the coronagraphs, we consider angular separations ($r$) from 0.5 $\lambda/D$ to 10 $\lambda/D$, and from 0.5 $\lambda/D$ to 2.5 $\lambda/D$ for the PLN.

\begin{figure}[H]
    \centering
    \includegraphics[width=0.7\linewidth]{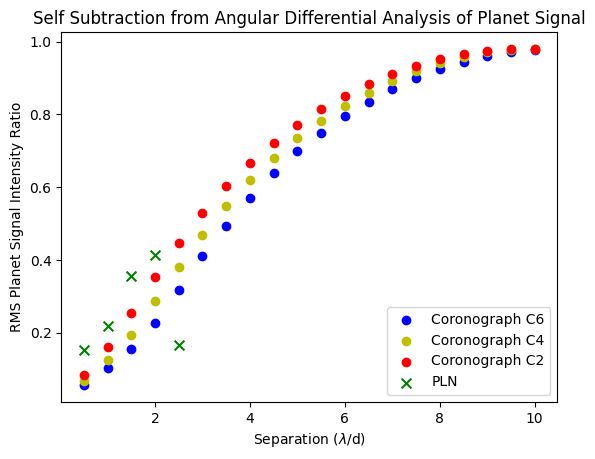}
    \caption{Fraction of a planet's RMS signal intensity retained after median subtraction modelling with itself, characterizing self-subtraction for Charge 2, 4, and 6 coronagraphs and a mode-selective PLN. The retained signal for the PLN drops sharply at $2.5~\lambda/D$, corresponding to a decrease in the instrument's throughput (see Fig.~\ref{fig:PLNInfo}) and indicating this separation is outside the PLN's effective field-of-view.}
    \label{fig:RMS_int}
\end{figure}

We start with an astrophysical planet signal ($A$) and calculate the self-subtracted flux ($F$) using median subtraction. For the coronagraphic signals, we define a circular planet signal window of radius 42 pixels ($W$) (capturing the ``useful throughput'' which includes $\sim 98\%$ of the total signal), centered at the pixel with the highest intensity. For the PLN, the window is simply all of the 4 nulled ports. Therefore, for practical algorithmic purposes, $W_{PLN} = [1,4]$, which refers to the 4 nulled ports out of 6 intensity points in the PLN. The ratio of RMS intensities in the windows' pixels/ports is then averaged over the frames to characterize their comparative self-subtraction effects at these separations (Fig. ~\ref{fig:RMS_int}).

\begin{align}
    F_{r, n} \;=\; A_{r, n}
\;-\;
\operatorname{median}\bigl\{A_{r, k}\colon k\neq n\bigr\} \\ 
\textrm{Ratio}(r) = \operatorname{mean}_{n}(\frac{\operatorname{RMS}(F_{r,n}[W_{r,n}])}{\operatorname{RMS}(A_{r,n}[W_{r,n}])}) 
\end{align}
where the indices $(r,n)$ refer to the angular separation index of the signal and the angular offset frame index, respectively.

The magnitude of the RMS ratio indicates the fraction of intensity remaining after self-subtraction. Therefore, a higher ratio corresponds to lower self-subtraction at that angular separation ($r$). Using this ratio as a metric, we find that the level of self-subtraction in a PLN at $1.0 \lambda/D$ is comparable to that of a charge 6 coronagraph at $2.0 \lambda/D$. Similar comparisons can be made across the entire throughput range of the PLN, which helps characterize the extent of self-subtraction that must be addressed during the post-processing of PLN data. As expected from the PLN's more rapid decrease in PSF correlation as a function of angular offset (shown in Fig. \ref{fig:Psf_dot_products}), for PSFs with the same angular sampling at the same separation, the PLN is less affected by self-subtraction than a coronagraph would be(until the separation exceeds the PLN's field- of- view). However, the overall level of self-subtraction remains high, and stellar subtraction methods for the PLN will have to account for this with some form of forward modeling.

This characterization of self-subtraction, (caused by overlapping planet signals at lower separations), the angular dependence of instrument signals, and the lack of physical significance in signal derotation and stacking all highlight the need to reformulate ADI-based post-processing algorithms to properly treat the data from a PLN. This method is also applicable to other instruments which produce non-rotationally invariant off-axis PSFs \cite{PorHaffert_2020, Haffert_2020, Blind_2025, Deshler_2025}.

\section{Methods}

\subsection{Generating Simulated Data}

To test and compare our post-processing algorithms, we first generate synthetic PLN datasets with correlated stellar noise. We generate phase aberrations in the pupil plane using the \lstinline{make_power_law_error} function from the \lstinline{hcipy} package (locally modified to accept a predetermined seed for random number generation, such that the phase screens can be reproduced). We first generate the phase screen $\Psi[0]$ with a peak-to-valley value of $0.3\pi$ and an exponent of -2.5. We propagate this initial phase screen $\Phi_0=\Psi_0$ through our model of the PLN to create the first frame of data with this initial aberration.

To evolve the phase aberration for each successive frame of the dataset, we first generate a new phase screen $\Psi[n]$ (with the same peak-to-valley and exponent parameters as above) for each frame, indexed by $n$. We then add a scaled version of this phase screen to the previous phase aberration, such that the phase aberration used to generate the data is

\begin{equation}
\label{eq:synth_gen}
    \Phi[n] = \sqrt{1-\zeta^2}\Phi[n-1] + \zeta \Psi[n].
\end{equation}

Here, $\zeta$ controls the rate of the drift, with higher values of $\zeta$ corresponding to more rapid evolution. We use $\zeta = 0.009$ for this work, corresponding to a slowly evolving quasistatic phase error. To isolate our analysis to the effect of ADI on quasistatic aberrations, we do not add any atmospheric turbulence. Our analysis is thus most applicable to the case in which the atmospheric turbulence averages out to a term that is static across all frames, and future work analyzing on-sky data (which has already been obtained) will be able to test the extent to which this generalizes.

To simulate companions that rotate in an arc around the central star, we generate a sequence of (unaberrated) off-axis PSFs corresponding to companions at different positions along the arc. We then add the sequence of companion PSFs to the stellar frames to create synthetic data of a binary system observed in pupil-tracking mode.

We initially generate 10000 datasets of 320 frames each; however, we choose to keep the angular rotation between successive frames (given by $\delta \theta$) fixed at 1.25 degrees, and the position angle of the companion in the initial frame (given by $\Theta$) fixed at 20 degrees. Therefore, depending on the overall amount of sky rotation we are modeling, we take a subset of the 320 frames for our desired synthetic dataset. For example, to test the algorithms for the case where the sky only rotates by 25 degrees total during the course of the observation (given by $\Delta \theta$), we take the subset of the first 20 frames of each dataset ($1.25^{\circ}\times 20=25^{\circ}$), and for the case where the sky rotates by 100 degrees total, we take the subset of the first 80 frames ($1.25^{\circ}\times 80=100^{\circ}$).

In theory, the sky rotation between successive frames depends on the exposure time of the detector. In practice, with typical short exposure times during which rotation within the exposure is negligible, it can be a parameter chosen in the data analysis step. For ADI with conventional coronagraphs, it is common to enforce an exclusion angle to minimize the correlation of noise as well as the overlap of the planet PSF between successive frames \cite{marois_adi}. However, enforcing such a condition is impractical at the spatial scales of $1 \lambda/D$ where the PLN operates, as choosing to avoid significant PSF overlap would require long observation times for significant rotation and also involve throwing out most of the data collected. We thus choose a small $\delta\theta$ between frames to approximate continuous rotation, such as if all the frames of the observation are included in the data analysis step. This has the advantage of maximizing the analysis to include all the data available.

\subsection{Estimation Algorithms} 
\label{sec: algos}

We introduce and compare three different algorithms for processing a PLN dataset taken in ADI (e.g. pupil-tracking) mode: one algorithm without any stellar subtraction (for comparison purposes), and two different algorithms for estimating and subtracting the stellar PSF. We denote the relevant astrophysical planet signal as $A$, which depends on the flux ratio ($\epsilon$), separation ($r$), and the on-sky angle of the initial frame ($\Theta$). These parameters define the cubic space over which the algorithms are evaluated. The algorithms assume the combined science intensity vector,
\begin{align}
    I_{j} = S_{j} + A_{j}(r, \epsilon, \Theta)
\end{align}
where the coordinate $j \in [1,4]$ refers to the four pixel intensities of the PLN signal, and $S$ represents the true stellar signal (under the effect of wavefront error). Algorithmically, we account for instrument sampling over the entire $\Delta \theta$ range, which determines the number of reference images used. We assume $\delta \theta = 1.25^{\circ}$, representing the angular offset between successive data frames taken at a sky rotation of $\delta \theta$. For all algorithms, the planet signal corresponding to the initial guess of the spatial parameters is denoted as $\hat{A}(\hat{r}, \hat{\epsilon}, \hat{\Theta})$. We describe the algorithms below. 

\subsubsection{Direct Fitting}
This algorithm fits the guessed planet signal ($\hat{A}(\hat{r}, \hat{\epsilon}, \hat{\Theta})$) to the input intensity ($I$), and calculates the cost as the mean of squared stellar residuals ($\hat{S}$).

\begin{align}
    \mathcal{J} =  \left\langle \left(I - \hat{A}(\hat{r}, \hat{\epsilon}, \hat{\Theta})\right)^2 \right\rangle = \left\langle \left(\hat{S}^2 \right)\right\rangle
\end{align}
where $\langle \cdot \rangle$ denotes the mean over all frames. 
This algorithm principally acts as a baseline for comparing the performance and statistical reliability of the other two algorithms.

\subsubsection{KLIP Subtraction}

This algorithm uses a modified version of  KLIP application discussed in Soummer et al. (2012) to process 1D data from the PLN. Estimating the stellar signal using KLIP subtraction for an input PLN intensity ($I$) and a guess planet ($\hat{A}(\hat{r}, \hat{\epsilon}, \hat{\Theta}$)) consists of the following steps. 

\begin{enumerate}
    \item For each frame of the input intensity, we compute a stellar estimate. To compute this estimate, we use all the frames (excluding the current one) as the reference frames ($R$). We start by decomposing the matrix of the reference frames $R$ using SVD.  
    \begin{align}
    R \;=\; U\,\Sigma\,\bigl(V\bigr)^{T}
    \end{align}
    \item Choose K right singular vectors with $K \in [1,4]$ limited by the (frames, 4) shape of each image for $I$ to form the component matrix $C_{k}$  
    \begin{align}
    C_{K} = V_{:,m},\; m=1,\dots,K
    \end{align}
    \item Compute the projection of the frame onto the span of K right singular eigenvectors (${v_{1}, ....v_{K}}$) of the reference matrix R. 
    \begin{align}
        \hat{S}[n] = C_{K}^{T} C_{K} I[n]
    \end{align}
    where $n$ refers to the frame index of the input intensity and the stellar estimate. 
    \item Forward model your planet signal by computing the scales associated with the KLIP components. 
    \begin{align}
    \vec{\alpha}[m] \;=\; \frac{\hat{A}[m]^{T}\,\hat{S}[m]}
    {\;\hat{I}[m]^{T}\,\hat{S}[m]\;}\\
    \hat{A}_{FM} (\hat{r}, \hat{\epsilon}, \hat{\Theta}) = \hat{A}(\hat{r}, \hat{\epsilon}, \hat{\Theta}) - Diag(\alpha)\hat{S}
    \end{align}
    where $\vec{\alpha}$ refers to the vector of relevant scale factors per frame, and $\hat{A}_{FM}$ is the forward modeled planet signal.
    \item Compute a final cost function using the calculated quantities.
    \begin{align}
    \label{eq:dir_fit_cost}
        \mathcal{J} =  \left\langle \left(I - \hat{S} - \hat{A}_{FM}(\hat{r}, \hat{\epsilon}, \hat{\Theta})\right)^2 \right\rangle
    \end{align}

\end{enumerate}

\subsubsection{Antiplanet KLIP Subtraction}
This algorithm is similar to the KLIP Subtraction method, but with the injection of a fake antiplanet to reduce self-subtraction effects before the stellar signal is estimated. This process is akin to the point-wise KLIP-FM method \cite{Pueyo_2016}, which involves a separate fake planet injection and KLIP reduction for every potential location of the planet. This process was deemed too computationally expensive to conduct over the entire field-of-view of a coronagraph, but is very feasible with an instrument like the PLN, which has much lower-dimensional data than that of a coronagraph, along with a much smaller field-of-view. The implementation of this algorithm can be described as follows. 

\begin{enumerate}    
    \item Calculate the guess-planet-subtracted input intensity.
    \begin{align}
        \hat{I}_{res} = I - \hat{A}(\hat{r}, \hat{\epsilon}, \hat{\Theta})
    \end{align}
    where $\hat{I}_{res}$ refers to the residual input intensity.
    \item Now, same as before, for each frame of the residual input intensity ($\hat{I}_{res}$) , decompose the matrix of the respective reference images $\hat{R}_{res}$ using SVD. And then choose the K right singular vectors with $K \in [1,4]$ to form the component matrix $\hat{C}_{k}$. 
    \begin{align}
        \hat{R} \;=\; \hat{U}\,\hat{\Sigma}\,\bigl(\hat{V}\bigr)^{T} \\
    \hat{C}_{K} = \hat{V}_{:,m},\; m=1,\dots,K
    \end{align}
    \item Compute the projection of the frame onto the span of K right singular eigenvectors (${\hat{v}_{1}, ....\hat{v}_{K}}$) of the reference matrix $\hat{R}$ to get your antiplanet stellar estimate. 
    \begin{align}
        \hat{S}_{anti}[n] = \hat{C}_{K}^{T} \hat{C}_{K} \hat{I}_{res}[n]
    \end{align} 
    \item Compute a similar looking final cost function using the antiplanet stellar estimate and the guess planet signal. 
    \begin{align}
        \mathcal{J} =  \left\langle \left(I - \hat{S}_{anti} - \hat{A}(\hat{r}, \hat{\epsilon}, \hat{\Theta})\right)^2 \right\rangle
    \end{align}
    Here we do not forward model the signal since the antiplanet injection before the stellar estimate is supposed to have taken care of the aggressive self-subtraction at these separations  ($r$). 

\end{enumerate}

We evaluate these algorithms over a cubic parameter space of $\hat{r}$, $\hat{\epsilon}$, and $\hat{\Theta}$ to characterize their spatial localization behavior for PLN images using different metrics. The confidence in the localizations from each method is compared using Monte Carlo-based detection testing of these algorithms. We also evaluate their performance for different $\Delta \theta$ ranges covered by the telescope pupil in the sky: $25^{\circ}$, $100^{\circ}$, and $150^{\circ}$. We then consider the pairwise joint localization behaviors for the three parameters $r$, $\epsilon$, and $\Theta$, and for the $\Delta \theta$ ranges, at the respective detection limits of each method. 

\section{Statistical Comparison}

\subsection{Detection Testing for Known Signal}
\label{sec:Det_test}
We perform statistical hypothesis tests to claim detections with each method and compare the results. We use the negative of the calculated cost from each method as our test statistic ($\mathcal{T}$), so that a higher cost indicates a worse fit for the astrophysical signal. Detection limits are determined using the Monte Carlo method, in which ten thousand random noise datasets are generated for stellar noise using the method described in Section 2.2. A PLN astrophysical planet signal ($A$) is then generated at a given flux ratio ($\epsilon$), for a specified $\Delta \theta$ range, at a known angular separation ($r$), initial angle $\Theta = 20^{\circ}$, and frames at $1.25^{\circ}/\text{frame}$. The alternative hypothesis ($H_{1}$) dataset includes the stellar noise signals combined with the known astrophysical planet signal, while the null hypothesis ($H_{0}$) dataset includes only the stellar noise signals.

\begin{align}
        \mathcal{T} =  -\mathcal{J}\\
        \mathcal{D}_{1, i} = I_{j} = S_{j} + A_{j}(r, \epsilon, \Theta)\\
        \mathcal{D}_{0, i} = I_{j} = S_{j}
\end{align}
With $i \in [1, 10000], j\in  [1,4]$. \\ 

The test statistics from each method are grouped into bins, and the frequency distributions of the test statistics from null datasets and alternative datasets are plotted for each bin. A smaller overlap between these two histograms for each method depicts higher confidence in detections for that method for the known planet signal ($A(r, \epsilon, \Theta)$). This overlap is quantified by treating each bin as a detection threshold ($\tau$), which defines the boundary above which the test statistic of an alternative hypothesis dataset for a given method ($\mathcal{T}(\mathcal{D}_{1, i})$) should fall (corresponding to a lower cost, $\mathcal{J}$), yielding the True Positive Rate (TPR) at that detection threshold for that method. Similarly, comparing it to the test statistic of a null hypothesis dataset for a given method ($\mathcal{T}(\mathcal{D}_{0, i})$), which also corresponds to a low cost ($\mathcal{J}$) for a dataset with no planet, yields the False Positive Rate (FPR) at that detection threshold for that method. Evaluating the TPR and FPR across all detection thresholds ($\tau$) enables plotting the receiver operating curves (ROC)  i.e., TPR vs. FPR   for a given method for the planet signal at these known parameters ($A(r, \epsilon, \Theta)$). The TPR and FPR are normalized over all datasets in the Monte Carlo analysis as shown below.
\begin{align}
    TPR(\tau) = \frac{1}{|\mathcal{D}_{1}|} \sum^{|\mathcal{D}_{1}|}_{i = 1} \textbf{I}[\mathcal{T}_{(\mathcal{D}_{1, i})} \geq \tau] \\
    FPR(\tau) = \frac{1}{|\mathcal{D}_{0}|} \sum^{|\mathcal{D}_{0}|}_{i = 1} \textbf{I}[\mathcal{T}_{(\mathcal{D}_{0, i})} \geq \tau]
\end{align}
Where \textbf{I} represents the indicator function \textbf{I} = \{1 if true, 0 if false\} \\  
We perform these detection tests for each method at the three different $\Delta \theta$ ranges mentioned earlier: $25^{\circ}$, $100^{\circ}$, and $150^{\circ}$. We make the known astrophysical planet signal ($A$) fainter or brighter by varying the Flux Ratio ($\epsilon$) to achieve $TPR = 0.9$ and $FPR = 10^{-3}$. The Flux Ratio ($\epsilon$) at which a method reaches these true- and false-positive rates is defined as the ``detection limit" ($\epsilon_{detlim}$) for that method, for the specified $\Delta \theta$ range and angular separation ($r$). For the KLIP subtraction and Antiplanet KLIP subtraction methods, these detection limits also depend on the number of right singular vectors ($K \in [1,4]$) selected for the stellar projection, denoted in the plots as $N_{comp}$. The value chosen is the one that yields the best performance for each method at the given spatial parameters in detection testing and localization (Section 3.2). Our simulations, including the phase screens described in Eq.~\ref{eq:synth_gen}, are implemented at a single wavelength (which we set to 1 in this work). In order to keep computational costs low and enable Monte Carlo statistical testing over a large number of synthetic datasets, we perform most of our analysis with such monochromatic data, for which there are only four nulled outputs. However, the PLN is able and intended to be used in a spectroscopically dispersed mode. In Appendix~\ref{app_sec: spec_dispersion}, we show an example of how the post-processing methods explored in this work are easily extendable to multi-wavelength PLN data, which has a higher data dimensionality, and for which we would expect a larger number of KLIP components to play a role. Additionally, a PLN with a larger number of modes would also have an increased data dimensionality. While PLNs with higher mode numbers than 6 provide limited additional off-axis throughput for exoplanets \cite{Xin_2022}, they may still be useful for interferometric imaging \cite{kim_coherent_imaging_2024} and spectroastrometric measurements \cite{kim_spectroastrometry_onsky_2025} of brighter sources, and would also benefit from KLIP-based calibration methods.

\begin{figure}[H]
    \centering
    \includegraphics[width=0.7\linewidth]{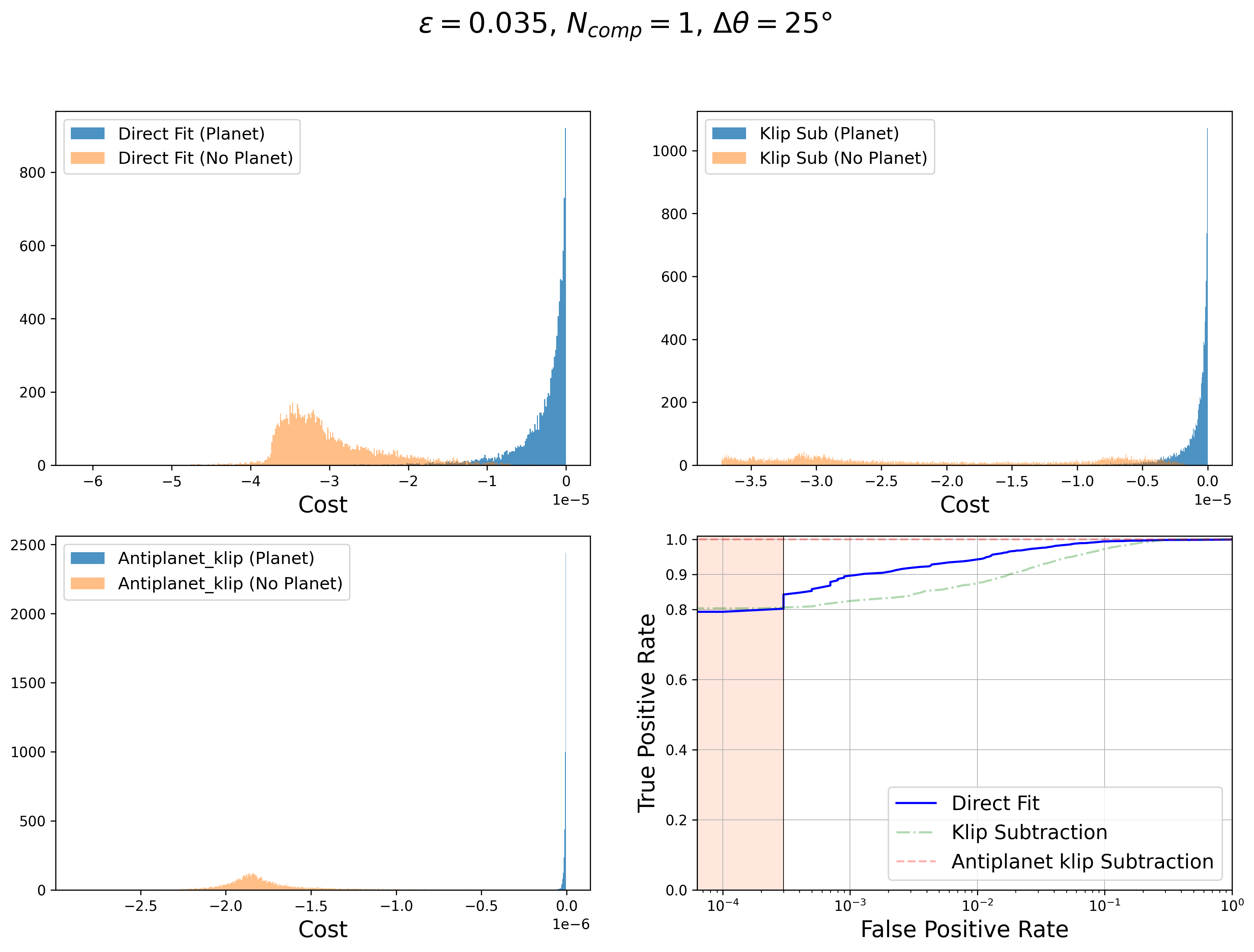}
    \caption{The detection limit plot for the Direct Fit method has 3 histograms and one ROC curve panel. The top-left histogram shows the test-statistic distributions for datasets with and without a planet for the Direct Fit method at its detection limit. The top-right and bottom-left histograms show the corresponding test-statistic distributions for the KLIP and Antiplanet KLIP methods, respectively, evaluated at this same flux ratio. Each histogram is focused on the region of overlap (if any) between the two distributions, which is the source of false positives. The bottom-right plot shows the corresponding ROC curves for each method. In the ROC plot, the shaded area highlights the region of poorly-sampled false positive rates, derived from fewer than 30 false detections. The ROC curve corresponding to the method for which this flux ratio is the detection limit (here, Direct Fit) is shown with higher opacity.}
    \label{fig:Complete_det_plot}
\end{figure}

\begin{figure}[H]
  \includegraphics[width = \textwidth]{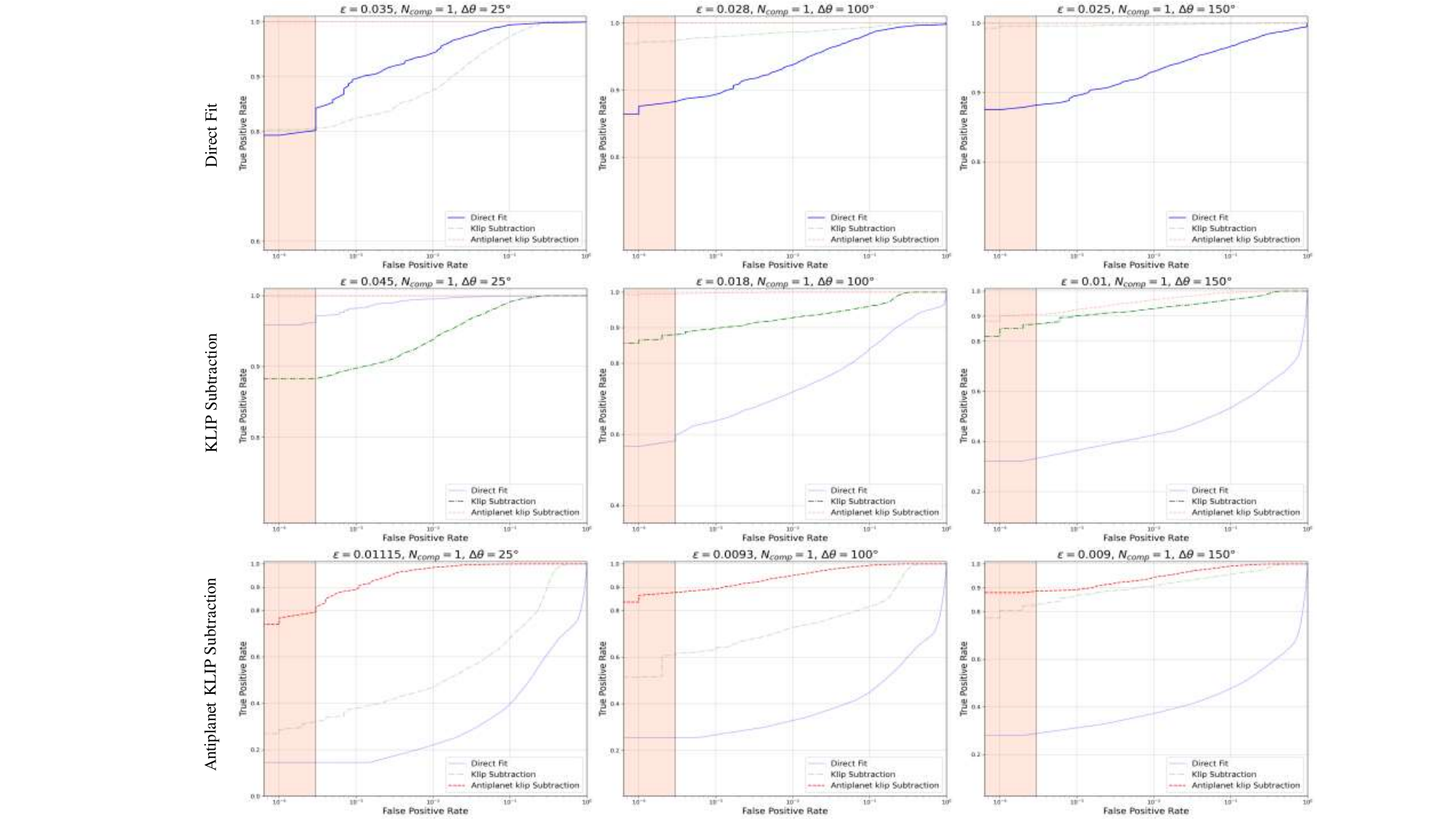}
  \caption{Detection limit ROC curves for the three localization methods, organized by reference method (rows) and $\Delta \theta$ (columns). Each row corresponds to the method for which the detection limit is evaluated, while each column shows results for $\Delta \theta = \{25^{\circ}, 100^{\circ}, 150^{\circ}\}$ from left to right. Within each panel, the ROC curves for all three methods are shown at the detection limit of the reference method. Row 1: Detection limit ROCs for Direct Fit (solid blue line) at its detection limit, compared with KLIP Subtraction (dashdot green line) and Antiplanet KLIP Subtraction (dashed red line). 
Row 2: Detection limit ROCs for KLIP Subtraction (dashdot green line) at its detection limit, compared with Direct Fit (solid blue line) and Antiplanet KLIP Subtraction (dashed red line). The detection limits for KLIP Subtraction are lower than Direct Fit at higher $\Delta \theta = \{100^{\circ}, 150^{\circ}\}$, and higher at $\Delta \theta = 25^{\circ}$. 
Row 3: Detection limit ROCs for Antiplanet KLIP Subtraction (dashed red line) at its detection limit, compared with Direct Fit (solid blue line) and KLIP Subtraction (dashdot green line). Given the correct astrophysical planet signal estimate, this method achieves the lowest detection limits for all $\Delta \theta = \{25^{\circ}, 100^{\circ}, 150^{\circ}\}$.}
  \label{fig:Det_Test_known_sig_direct}
\end{figure}

To keep our detection test comparisons consistent, the known planet signal for each method was spatially restricted to $r = 0.75 \lambda/D$ and $\Theta = 20^{\circ}$. The resulting ROC curves from these detection tests with a known planet signal are shown in Fig.~\ref{fig:Det_Test_known_sig_direct}. Fig.~\ref{fig:Complete_det_plot} presents a sample detection plot with test-statistic histograms illustrating the $D_{0}$ and $D_{1}$ overlap regions and their impact on the ROCs. The key takeaway is that the Antiplanet KLIP Subtraction method outperforms the other two methods by achieving a lower detection limit at a fixed angular separation ($r$) for all three $\Delta \theta$ ranges when the astrophysical planet signal is known. \\

The detection ROCs also vary minimally with a change in separation ($r$). Reducing the angular separation $r$ (while keeping $\Theta$ and $\epsilon$ fixed) decreases overall detection performance across all methods by reducing the TPR, whereas increasing $r$ improves it. However, these changes in ROC curves do not alter the relative performance of the three methods: the Antiplanet KLIP Subtraction continues to outperform the others, though the absolute detection significance varies with $r$. This behavior is consistent with expectations — fainter planets are more difficult to localize when they lie closer to the host star and easier to localize when farther away.

These detection limits for each localization algorithm at its corresponding $\Delta \theta$ are tabulated in Table ~\ref{tab:detlims}.

\begin{table}[h]
  \centering
  \begin{tabular}{|c|ccc|}
  \hline
   Localization Method 
    & $25^{\circ}$ 
    & $100^{\circ}$ 
    & $150^{\circ}$ \\
    \hline
    Direct Fit 
    & 0.035 
    & 0.028 
    & 0.025 \\
    KLIP Subtraction 
    & 0.045 
    & 0.018 
    & 0.01 \\
    
    Antiplanet KLIP Subtraction 
    & 0.01115 
    & 0.0093 
    & 0.009 \\
    \hline
  \end{tabular}
  \caption{Detection Limit flux ratio ($\epsilon_{detlim}$) for each method and their corresponding $\Delta \theta$ as tabulated from the ROC's in Fig. ~\ref{fig:Det_Test_known_sig_direct}.}
  \label{tab:detlims}
\end{table}

\subsection{Localization} \label{sec:localization}
To assess the spatial localization capability of the PLN after processing the data with our algorithms, we insert astrophysical signals ($A$) with known spatial parameters ($r, \epsilon, \Theta$) into stellar noise datasets and evaluate the costs of each algorithm for those datasets over a given parameter space. The flux ratios ($\epsilon$) for these signals are set to the detection limits for each respective method, while maintaining $r = 0.75 \lambda/D$. 
This localization is done for all three angular coverage ($\Delta \theta$) values $(25^{\circ}, 100^{\circ}, 150^{\circ})$, with each algorithm used to evaluate the dataset containing astrophysical spatial parameters at the detection limit of every other algorithm for the corresponding $\Delta \theta$.\\
We use 10 stellar noise datasets for each astrophysical planet signal at a given set of spatial parameters to obtain a distribution of the cost function's performance for each method, corresponding to every other method's detection limit ($\epsilon_{detlim}$) at a given $r, \Theta$. The number of datasets is limited to 10 by the computational time of evaluating three-dimensional parameter grids for each dataset, which becomes increasingly intensive with larger $\Delta \theta$ coverages. The final cost grids obtained are four-dimensional, with an additional dimension for the datasets. These grid sizes vary for different methods and tested parameters, depending on the number of test flux ratios ($\widehat{\epsilon}$) chosen relative to the known detection limit flux ratio {$\epsilon_{detlim}$} of the input planet. The grid size along the angular separation ($r$) and initial angle ($\Theta$) axes remains fixed, since our test arrays $\widehat{r}$ and $\widehat{\Theta}$ span $\widehat{r} \in [0, 2] \lambda/D$ at equally spaced intervals of $0.25\lambda/D$ and $\widehat{\Theta} \in [0, \pi]$ at equally spaced intervals of $\pi/18$. We formulate an unnormalized probability $\mathcal{P}$ distribution based on our cost values as in Eq. ~\ref{eq: localization_prob}. We then calculate the marginalized and joint distributions for each of the spatial parameters ($r$, $\epsilon$, $\Theta$) and their pairs. Here, the cost $\mathcal{J}$ is the corresponding quantity defined in Eq. 5, 11, and 16. 

\begin{align}
\label{eq: localization_prob}
        \mathcal{P}(\widehat{r}, \widehat{\epsilon}, \widehat{\Theta})_{d} \propto e^{-\frac{\mathcal{J}_{d, \widehat{r}[a], \widehat{\epsilon}[b], \widehat{\Theta}[c]}}{2}}   
\end{align}
With $d \in [1, 10], a\in  [1,9], b\in  [1,len(\widehat{\epsilon})]$ and $c\in [1,19]$. Where $I(r, \Theta, \epsilon)_{d}$ represents the (1,4) shaped intensity vector with the stellar signal dataset $d$ and the ``correct" planet at spatial parameters $(r, \epsilon, \Theta)$. ($\widehat{r}, \widehat{\epsilon}, \widehat{\Theta}$) are the test arrays for the $(r, \epsilon, \Theta)$ parameters corresponding to cost grid axes (1,2,3) respectively. $\mathcal{P}$ represents the joint distribution across the 3 parameters, $\mathrm{P}$ represents the joint distribution for a pair of parameters and $\mathbf{P}$ represents the marginalized distribution for a single parameter.\\

\begin{figure}[H]
  \centering
    \centering
    \includegraphics[width=\linewidth]{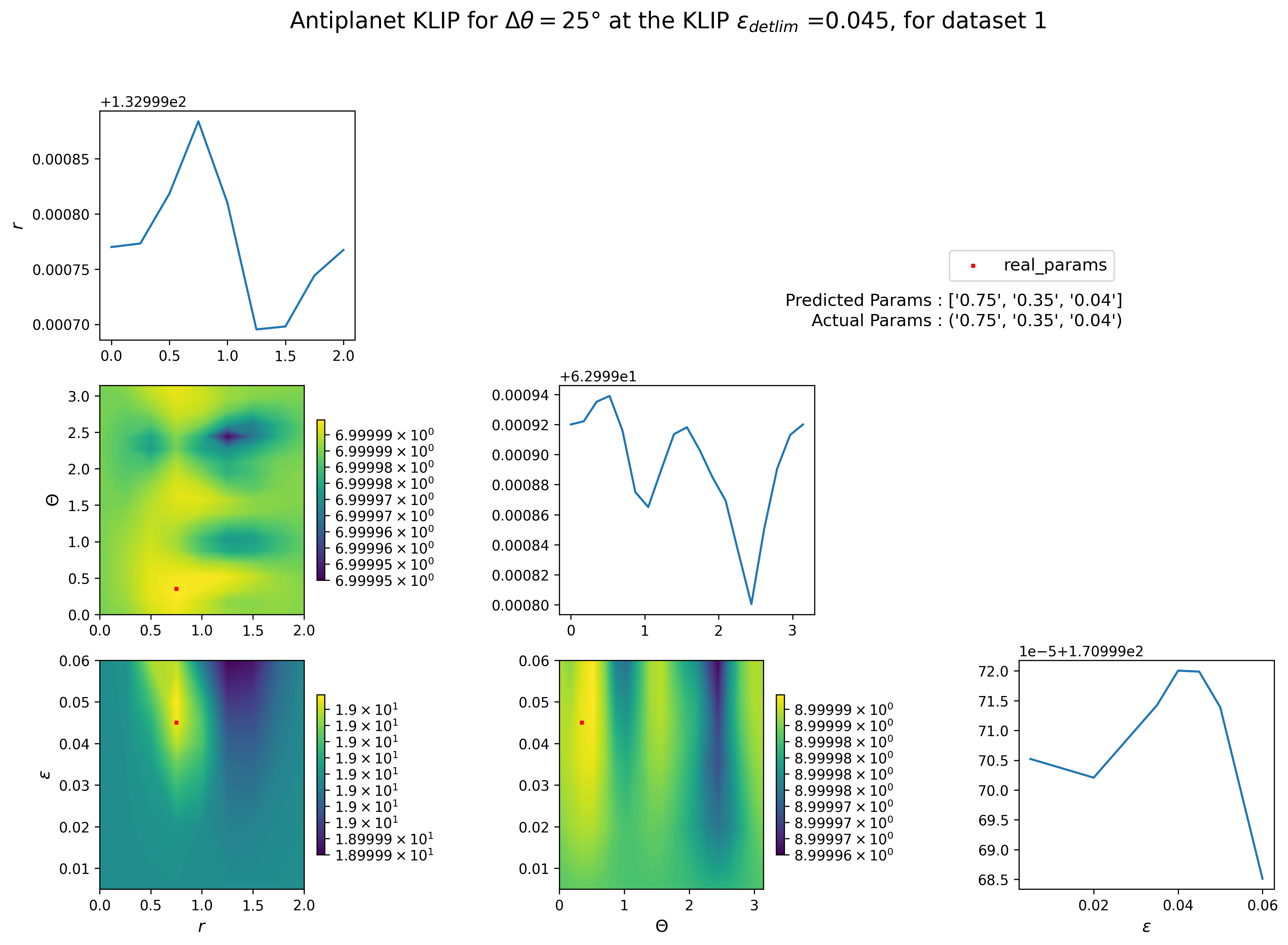}

    \caption{The localization corner plot depicts the accurate localization by Antiplanet KLIP subtraction at the KLIP subtraction detection limit ($\epsilon_{KLIP}$) for the dataset indexed 1. The diagonals of the corner plot show the marginal probability distributions of the parameter localizations, and the joint plots represent the pairwise probability distributions for the three parameters ($r, \Theta, \epsilon$). ``Predicted Params" tuple notes down the parameters with the highest probability (lowest cost) from the Antiplanet KLIP subtraction cost grid after localization.}

    \label{fig:localization_example}
\end{figure}

An example of the resulting localization triangle plot for the Antiplanet KLIP Subtraction method evaluated at the KLIP subtraction detection limit ($\epsilon_{\mathrm{KLIP}}$), is shown in Fig.~\ref{fig:localization_example} for a representative dataset. These plots illustrate how the cost function–based localization behaves for a single noise realization. Results for all the methods localizing at the KLIP subtraction detection limits at $\Delta \theta = 25^{\circ}$ and $100^{\circ}$ are summarized in Table.~\ref{tab:klip_detlim_localization_25_100} for comparison. A complete set of localization corner plots, including results for $\Delta \theta = 150^{\circ}$ and for detection limits corresponding to Antiplanet KLIP subtraction and Direct Fit, is provided in Appendix A. Overall, our results demonstrate that the Antiplanet KLIP subtraction method offers the most consistent spatial localization across $\Delta \theta$ coverages ($25^{\circ}$, $100^{\circ}$, and $150^{\circ}$). Specifically, the method achieves perfect localization across all datasets at $\Delta \theta = 100^{\circ}$ and $150^{\circ}$ for all detection-limit flux ratios. At $\Delta \theta = 25^{\circ}$, Antiplanet KLIP subtraction achieves perfect localization at the KLIP and Direct Fit detection limits, while correctly localizing 5 out of 10 datasets at its own, more challenging detection limit. Even in this more difficult regime, its localization performance exceeds that of both KLIP subtraction and Direct Fit on the same datasets and parameter grids.

\begin{table}[h]
  \centering
  \begin{tabular}{|c|c|c|c|c|}
    \hline
    Localization Method & KLIP Detection Limit & $\Delta \theta$ & ($r, \epsilon, \Theta$)& ($\tilde{r}, \tilde{\epsilon}, \tilde{\Theta}$)\\
    \hline
    \rule{0pt}{3ex} & & & & \\
    Antiplanet KLIP Subtraction & 0.045 & $25^{\circ}$ & (0.75, 0.045, $20^{\circ}$)& (0.75, 0.045, $20^{\circ}$)\\
    KLIP Subtraction & 0.045 & $25^{\circ}$ & (0.75, 0.045, $20^{\circ}$)& (0.75, 0.04, $20^{\circ}$)\\
    Direct Fit & 0.045 & $25^{\circ}$ & (0.75, 0.045, $20^{\circ}$)& (0.75, 0.05, $20^{\circ}$)\\
    Antiplanet KLIP Subtraction & 0.018 & $100^{\circ}$ & (0.75, 0.018, $20^{\circ}$)& (0.75, 0.018, $20^{\circ}$)\\
    KLIP Subtraction & 0.018 & $100^{\circ}$ & (0.75, 0.018, $20^{\circ}$)& (0.75, 0.018, $20^{\circ}$)\\
    Direct Fit & 0.018 & $100^{\circ}$ & (0.75, 0.018, $20^{\circ}$)& (0.75, 0.025, $20^{\circ}$)\\
    \hline
  \end{tabular}
  \caption{Localization at the KLIP subtraction detection limits for $\Delta \theta =  \{25^{\circ}, 100^{\circ}\}$ with Astrophysical signal parameters, $A(r, \Theta) = 0.75, 20^{\circ}$. $(\tilde{r}, \tilde{\epsilon}, \tilde{\Theta})$ represents the parameters with the maximum likelihood.}
  \label{tab:klip_detlim_localization_25_100}
\end{table}

\subsection{Detection Testing For Robustness}
\begin{figure}[H]
  \centering
  \includegraphics[width = \linewidth]{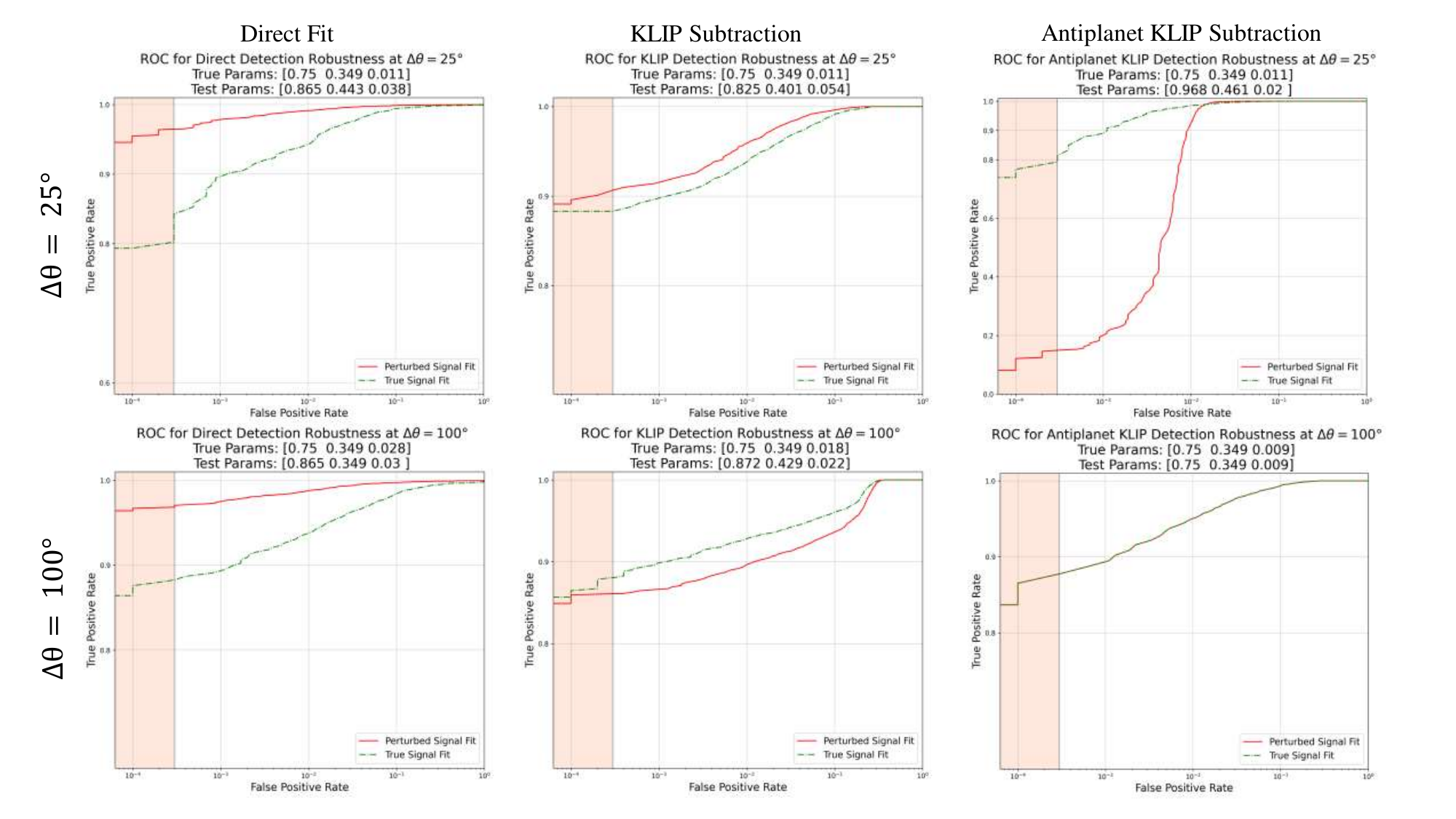}
  
  \caption{Detection testing with imperfect planet localization. The panels are organized by $\Delta \theta$ (rows) and localization method (columns).  Perturbed input parameters are given as ``Test Params" in the plot. The ``True Params" for the planet signal $A(r, \Theta, \epsilon)$ in $\mathcal{D}_{1}$ datasets are also shown. The perturbed input parameters come from adding the standard deviation of the localization errors at the corresponding $\Delta \theta$, resulting in the modeled planet being farther from the star than the true planet. Top row ($\Delta \theta = 25^{\circ}$): Direct Fit and KLIP ROCs achieve a higher TPR at the corresponding FPR, while the Antiplanet KLIP ROC worsens. Bottom row ($\Delta \theta = 100^{\circ}$): Direct Fit ROC achieves a higher TPR at the corresponding FPR, while the KLIP ROC worsens. Antiplanet KLIP ROC stays the same due to no localization errors.}
  \label{fig:robust_direct_klip_pos}
\end{figure}

\begin{figure}[H]
  \centering
  \begin{subfigure}[t]{0.5\textwidth}
    \centering
    \includegraphics[width=\linewidth]{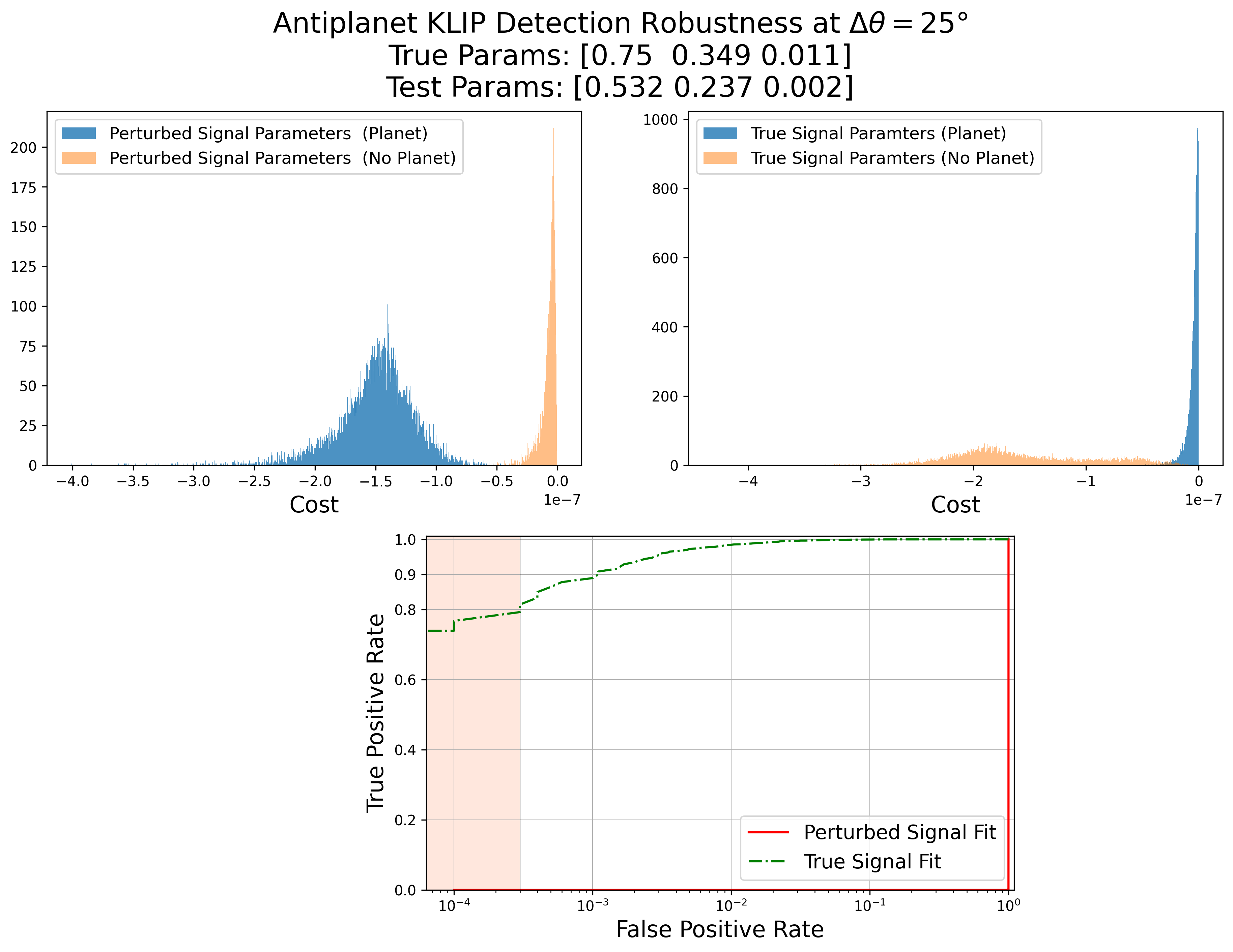}
    \subcaption{}\label{subfig:img1}
  \end{subfigure}\hfill
  \begin{subfigure}[t]{0.38\textwidth}
    \centering
    \includegraphics[width=\linewidth]{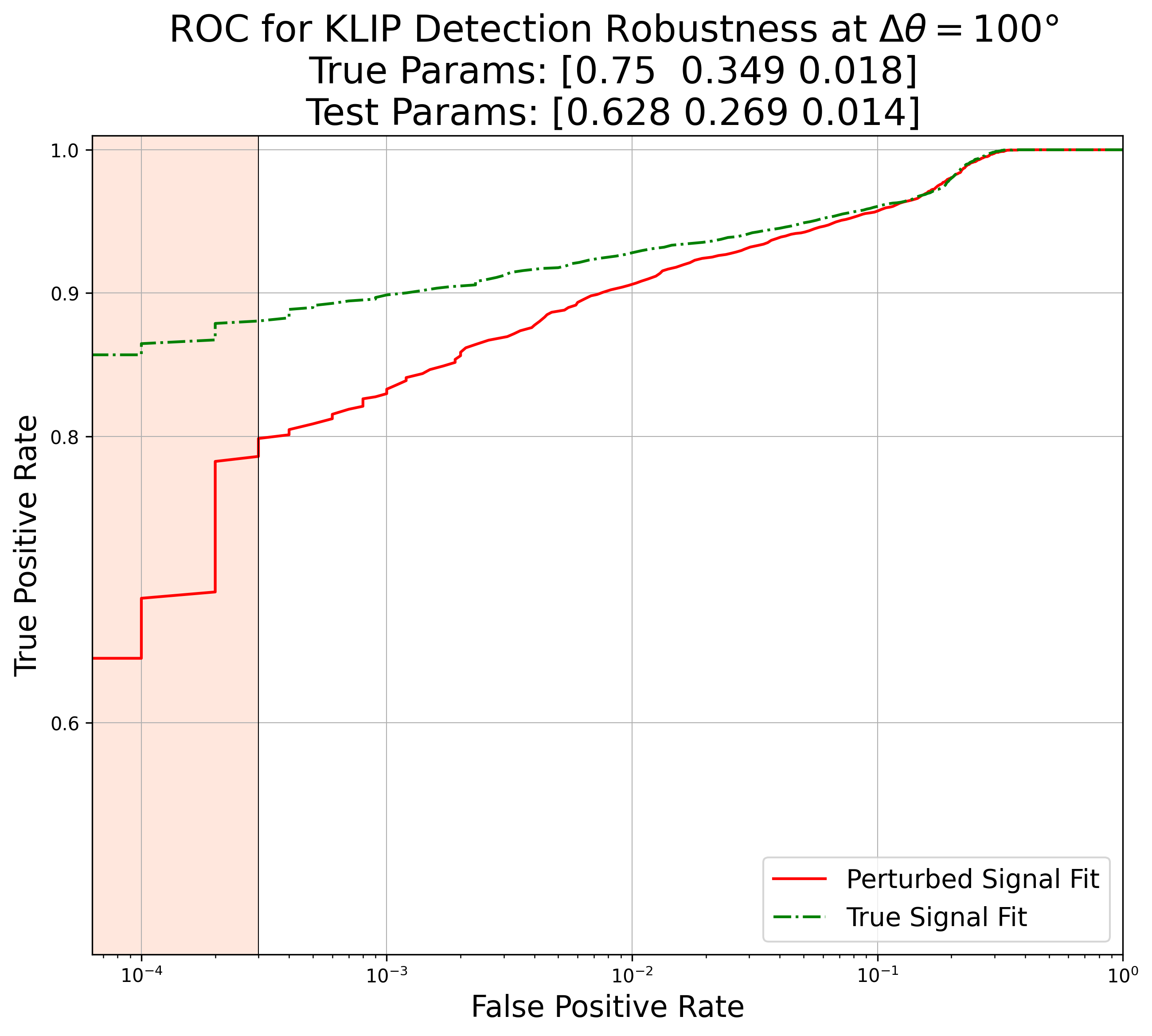}
    \subcaption{}\label{subfig:img2}
  \end{subfigure} \hfill
\caption{Detection testing with imperfect planet localization when the modeled planet is closer to the star than the true planet. The perturbed input parameters come from subtracting the standard deviation of the localization errors at the corresponding $\Delta \theta$. (a) The ROC indicates a ``bad fit" for these parameters, due to the low detection limit for Antiplanet KLIP for $\Delta \theta = 25^{\circ}$. Therefore, in the test-statistic histogram in the corresponding detection plot,  $\mathcal{D}_{1}$ datasets have a more negative cost than the $\mathcal{D}_{0}$ datasets. (b) The KLIP subtraction ROC worsens as it achieves a lower TPR at the corresponding FPR for $\Delta \theta = 100^{\circ}$.}
\label{fig:robust_anti_klip}
\end{figure}

Having established the localization performance of each method, we assess how robust their detection statistics are to localization uncertainty. For each method at its own detection limit, we quantify the spread in recovered spatial parameters across the 10 stellar noise datasets. The standard deviation of the minimum-cost ($\mathcal{J}_{min}$) localization ($r_{min}, \epsilon_{min}, \theta_{min}$) is treated as a perturbation to the true input planet signal, $\widehat{A}$.
\\
We then use this perturbed signal in place of the perfectly known input signal to compute cost-based test statistics (Eq.~16) for the alternative hypothesis datasets ($\mathcal{D}{1,i}$), \emph{which still contain the true injected planet} $A(r,\epsilon_{\mathrm{detlim}}, \theta)$. By comparing the resulting ROC curves to those obtained in the earlier detection limit tests, we directly see how imperfect signal localization influences the detection testing performance of each method across the three $\Delta \theta$ angles ($25^{\circ}$, $100^{\circ}$, $150^{\circ}$).  The ROC curves from this test are shown for the addition of positive localization error deviations  at $\Delta \theta = \{25^{\circ}, 100^{\circ}\}$, in Fig.~\ref{fig:robust_direct_klip_pos}. \\
The robustness tests for the Antiplanet KLIP method differ from the other two methods for two primary reasons. First, the Antiplanet KLIP is capable of detecting fainter planets than the other methods. The faint signals in the regime of the detection limit, when combined with limited angular coverage (in this case, $25^{\circ}$), cause imperfect localizability of the planet signal, to which the Antiplanet KLIP method is rather sensitive. Even minor perturbations from the true parameters result in a ``bad fit" of the data to the model, making the detection limit of this method not robust to deviations in signal localization. The ``bad fits" are well illustrated in the detection-testing histograms: not only do the test statistics from datasets with and without a planet have significant overlap, but the perturbation dominates the planet signal to the extent that the costs for the datasets without an astrophysical signal are lower than the costs for datasets with the perturbed signal. The Antiplanet KLIP-specific detection plot for $\Delta \theta = 25^{\circ}$ is shown in Fig.~\ref {fig:robust_anti_klip}. Second, the localization using Antiplanet KLIP with larger angular coverages ($100 ^{\circ}$ and $150^{\circ}$) shows that the maximum-likelihood parameters perfectly coincide with the true values for all datasets tested. For lower sky coverage ($\Delta \theta = 25^{\circ}$), although the KLIP subtraction method does not achieve as deep a detection limit as the Antiplanet KLIP method can for a perfectly known planet signal, it is more robust to the slightly inaccurate planet parameter estimates that result from the limited angular coverage of the dataset. Ultimately, although the detection limit for Antiplanet KLIP is the most sensitive to imperfect localizability, the method still outperforms KLIP and direct fitting at all $\epsilon$ for both detection and localization. \\

\section{Conclusion}
PLN is a novel instrument concept designed to image faint, close-in off-axis exoplanets at angular separations of $0.5$–$1.5$ $\lambda/D$. To fully exploit the angular diversity of pupil-tracking observations and reduce degeneracies in estimating the planet’s position, we reformulated the angular differential imaging (ADI) algorithm for use with PLN data. This generalized formulation can also be applied to other non-image-forming instruments lacking rotational invariance.

We quantified the self-subtraction effects introduced by ADI-like processing at small inner working angles for the PLN signal, which motivated the development of the Antiplanet KLIP Subtraction method. This approach, conceptually similar to the point-wise KLIP forward-modeling (point-wise KLIP-FM) method, is computationally efficient because of PLN’s lower data dimensionality.

Monte Carlo–based statistical detection tests demonstrate that the Antiplanet KLIP method outperforms both the Direct Fit baseline and the PLN-optimized KLIP implementation, detecting fainter planets with higher confidence using a single singular component for the stellar model across three sky coverages ($\Delta \Theta = {25^{\circ}, 100^{\circ}, 150^{\circ}}$). Across 10 representative datasets, estimates of the planet's parameters (separation $r$, flux ratio $\epsilon$, and initial frame angle $\Theta$) are the most accurate using the Antiplanet KLIP subtraction method.

However, when localization uncertainties are propagated into the detection tests, the detection limit of the Antiplanet KLIP method shows sensitivity to localization errors at smaller sky coverages ($\Delta \theta = 25^{\circ}$). This is because it is capable of probing fainter signals than the other methods, and thus suffers more from the uncertain localization given the limited angular coverage. Nevertheless, at $\Delta \theta = 25^{\circ}$, it remains the most robust method for detecting brighter planets also probed by the other methods (e.g. with flux ratios $\epsilon_{KLIP,\mathrm{detlim}}$ and $\epsilon_{Direct,\mathrm{detlim}}$), and is both sensitive and robust for data with higher angular coverages of  $\Delta \theta = {100^{\circ}, 150^{\circ}}$. We recommend maximizing sky coverage whenever possible. 

Overall, Antiplanet-KLIP is the best method to constrain a planet's on-sky position for the PLN due to its ability to probe the faintest signals and having the highest localization accuracy at all $\Delta \theta$. Future work will explore incorporating multi-wavelength data to extend the angular differential framework spectrally, potentially enabling improved sensitivity from spectrally disentangling residual starlight from planet light, as well as the ability to characterize the planet spectrally.

\subsection*{Disclosures}
The authors declare that there are no financial interests, commercial affiliations, or other potential conflicts of interest that could have influenced the objectivity of this research or the writing of this paper.

\subsection* {Code, Data, and Materials Availability} 
The code, data, and simulations used in this work can be found at \url{https://github.com/SuviIlli/pln-post_processing-jatis}, or with the DOI: 10.5281/zenodo.20835372.

\subsection* {Acknowledgments}

This work was supported by the National Science Foundation under Grant No. 2308360. We thank the California Institute of Technology for providing the SURF Fellowship that was crucial to the completion of this study. This research made use of HCIpy \cite{por2018hcipy}; Astropy \cite{astropy:2013,astropy:2018,astropy:2022}; NumPy \cite{harris2020array}; SciPy \cite{2020SciPy-NMeth}; and Matplotlib \cite{Hunter:2007_matplotlib}.
We acknowledge the use of a Grammarly browser extension for grammatical and typographical help.

\newpage
\bibliography{report}  
\bibliographystyle{spiejour}   

\newpage
\appendix
\section{Post-Processing the Data from a Spectrally Dispersed PLN}
\label{app_sec: spec_dispersion}

To simulate the operational capabilities of these algorithms when the photonic lantern is used in a spectrally dispersed mode on-sky, we also generate datasets with explicit wavelength dependence. Our datasets are generated over wavelengths in normalized units centered at $\lambda=1.0$ and spanning $0.9–1.1$ with a spacing of $0.02$, resulting in 100 measured wavelengths for each port. For simplicity, we have assumed that the flux of the star is constant across wavelengths, and that the flux ratio of the planet is also constant across wavelengths. The time-evolution of the phase error at the central wavelength ($\Phi[n, \lambda_0]$) is described by Equation \ref{eq:synth_gen}. In order to generate the corresponding phase error at nearby wavelengths, we assume that the phase error is due to a physical optical path difference that is the same across all wavelengths. We thus scale the phase screens for different wavelengths using $\Phi[n, \lambda]= (\lambda_0/\lambda)\Phi[n, \lambda_0]$. To simulate data formation through the lantern, we also assume that the mode structure of the lantern is constant across the wavelength range.

The synthetic science dataset is then formed by generating unaberrated off-axis PSFs corresponding to companions at different positions along the sky-rotation arc. Because the planet is at a fixed angular separation in radians, we scale the planet separation (when expressed in $\lambda/D$ units) by $\lambda^{-1}$ for each $\lambda$, before generating the corresponding PSF and adding it into the synthetic data. The analysis and implementation of the algorithms (Section~\ref{sec: algos}) for these datasets are then performed by stacking the nulled ports across all wavelengths, forming a data vector whose size is $N_{\mathrm{ports}}N_{\mathrm{wavels}}$.

The example analysis is performed for $\Delta \theta = 100^{\circ}$ while maintaining $\delta\theta = 1.25^{\circ}$ and $\Theta = 20^{\circ}$. This results in 80 frames in the datasets, with the additional spectral dimension yielding a dataset of shape $(80, 100, 4)$ for the four nulled PLN ports. The nulled-port intensities are then stacked across all wavelength modes, producing an overall dataset of shape $(80, 400)$.

\begin{figure}[H]
\centering
\includegraphics[width=1\linewidth]{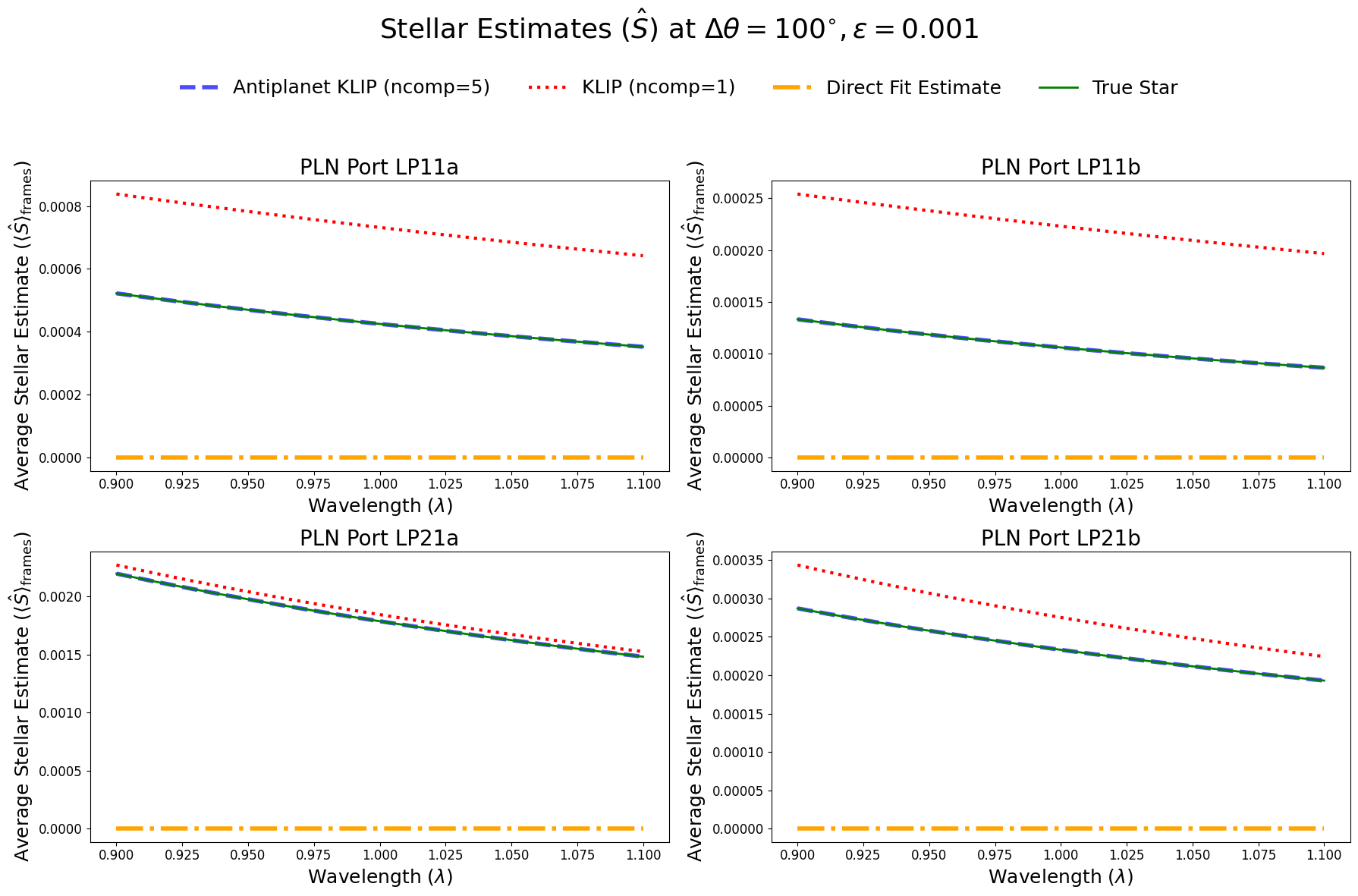}
\caption{Stellar estimates from localization algorithms for each of the four nulled PLN ports, averaged over all 80 spatial frames at $\Delta \theta = 100^{\circ}$ for a spectrally dispersed dataset. Antiplanet KLIP with (with optimal $ncomp = 5$) most accurately recovers the true stellar signal across all ports with flux ratio $\epsilon = 0.001$. In contrast, KLIP (with optimal $ncomp=1$) produces a less accurate estimate of the stellar signal in this faint regime. The Direct Fit baseline yields a zero-vector stellar estimate, reflecting the absence of an explicit stellar model in its cost function. The lower flux ratios retrievable relative to the corresponding monochromatic dataset highlight the improved sensitivity due to additional wavelength information.}
\label{fig:multiwave_star}
\end{figure}

Fig.~\ref{fig:multiwave_star} compares the average stellar estimates across spatial frames from these algorithms for one representative high-dimensional multiwavelength dataset. These stellar estimates are obtained by evaluating the algorithms on the combined science intensity vector $I$, containing the companion signal $A(r(\lambda), \epsilon = 0.001, \Theta = 20^{\circ})$, and using $\hat{A}(\hat{r}(\lambda), \hat{\epsilon}, \hat{\Theta}) = A(r(\lambda), \epsilon, \Theta)$ for Antiplanet KLIP. The quantity $\langle \hat{S} \rangle_{\text{frames}}$ is taken to be a zero vector for the Direct Fit method, since the cost function in Eq.~\ref{eq:dir_fit_cost} does not include an explicit stellar estimate.

We find that $ncomp = 5$ outperforms $ncomp=1$ for Antiplanet KLIP, reflecting the increased dimensionality of the WFE, where projecting onto a larger number of components improves the estimate for the stellar signal, whereas using a single component yields a less accurate stellar estimate. In contrast, KLIP remains optimal with a single component as using higher components removes a significant fraction of the planet signal, an effect related to self-subtraction. The agreement of Antiplanet KLIP with the true stellar signal across all PLN ports at $\epsilon = 0.001$ demonstrates its accuracy in this higher-dimensional spectrally dispersed setting. 

Furthermore, localization analysis of five representative datasets over a grid of parameters (as in Section \ref{sec:localization}), using the three reduction methods at different flux ratios also shows that incorporating wavelength information enables the retrieval of fainter planets than was possible using monochromatic data.

\section{Localization for $\Delta \theta = {100^{\circ}, 150^{\circ}}$}

\begin{figure}[H]
\centering
\includegraphics[width=0.75\linewidth]{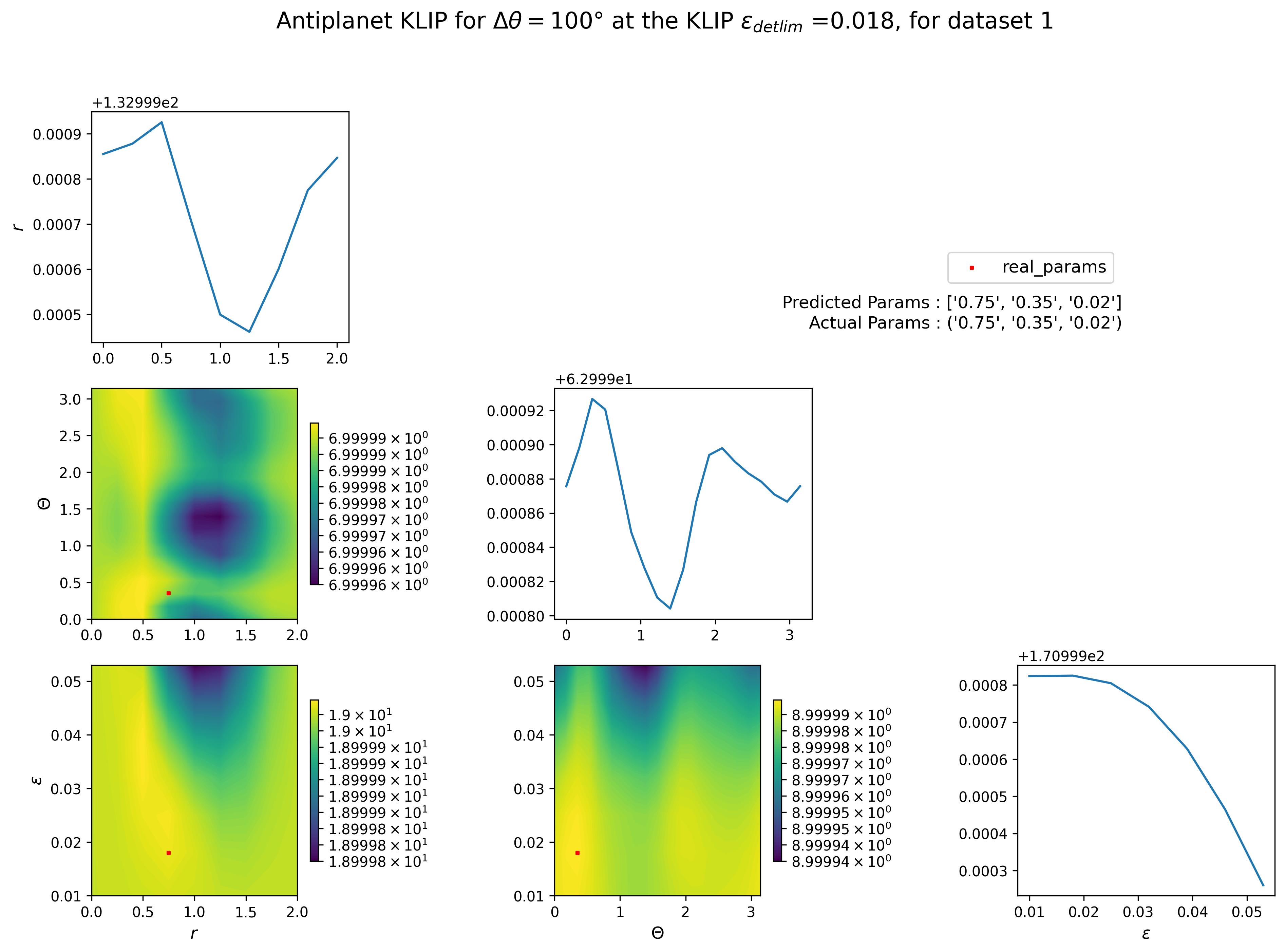}
\caption{Accurate localization achieved by the Antiplanet KLIP subtraction at $\Delta \theta = 100^{\circ}$ and $\epsilon_{KLIP}$ for dataset 1.}
\label{fig:100 deg localization example}
\end{figure}

\begin{figure}[H]
\centering
\includegraphics[width=0.75\linewidth]{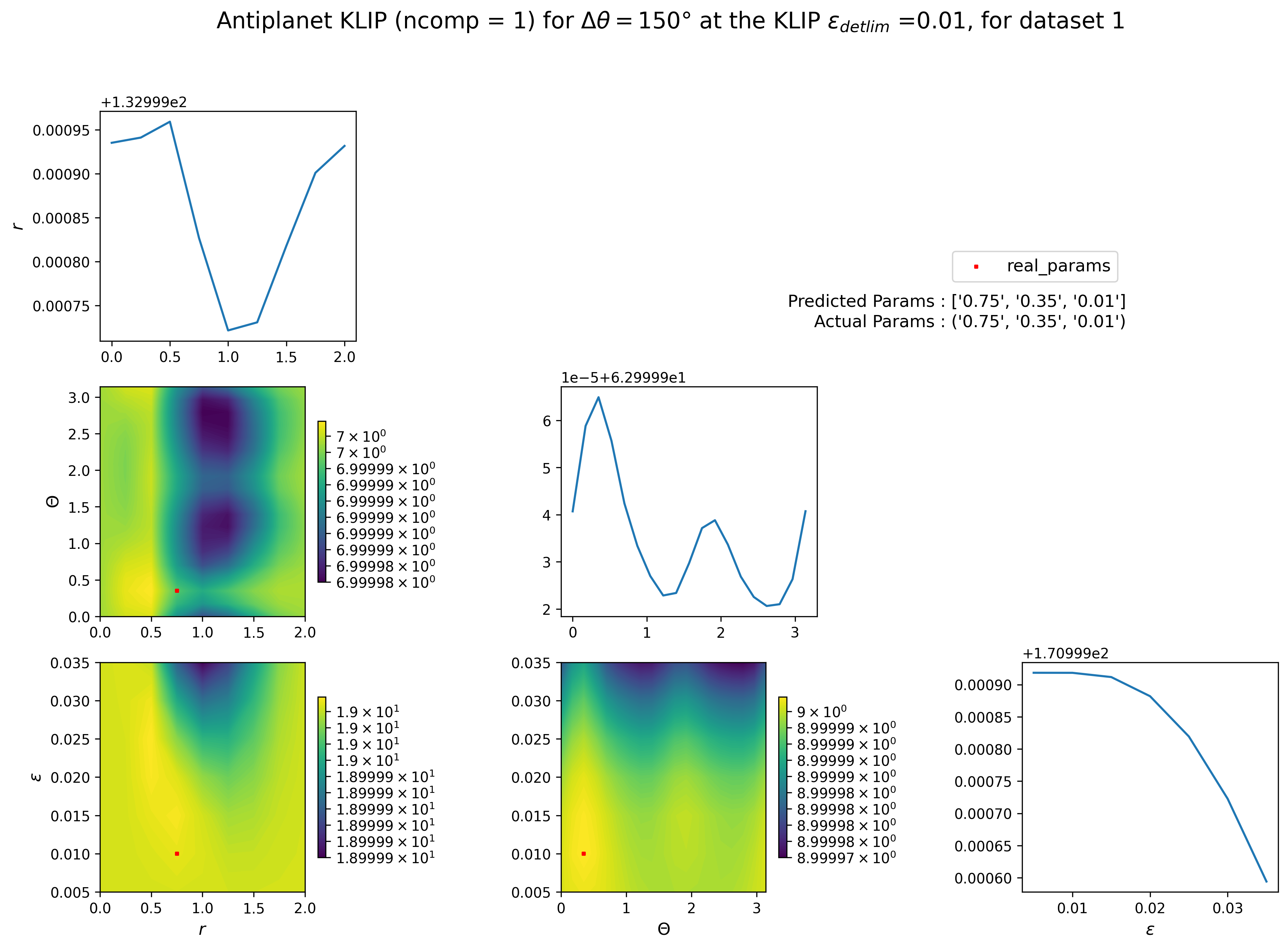}
\caption{Accurate localization achieved by the Antiplanet KLIP subtraction at $\Delta \theta = 150^{\circ}$ and $\epsilon_{KLIP}$ for dataset 1.}
\label{fig:Appendix KLIP 150}
\end{figure}

\begin{figure}[H]
\centering
\includegraphics[width=0.75\linewidth]{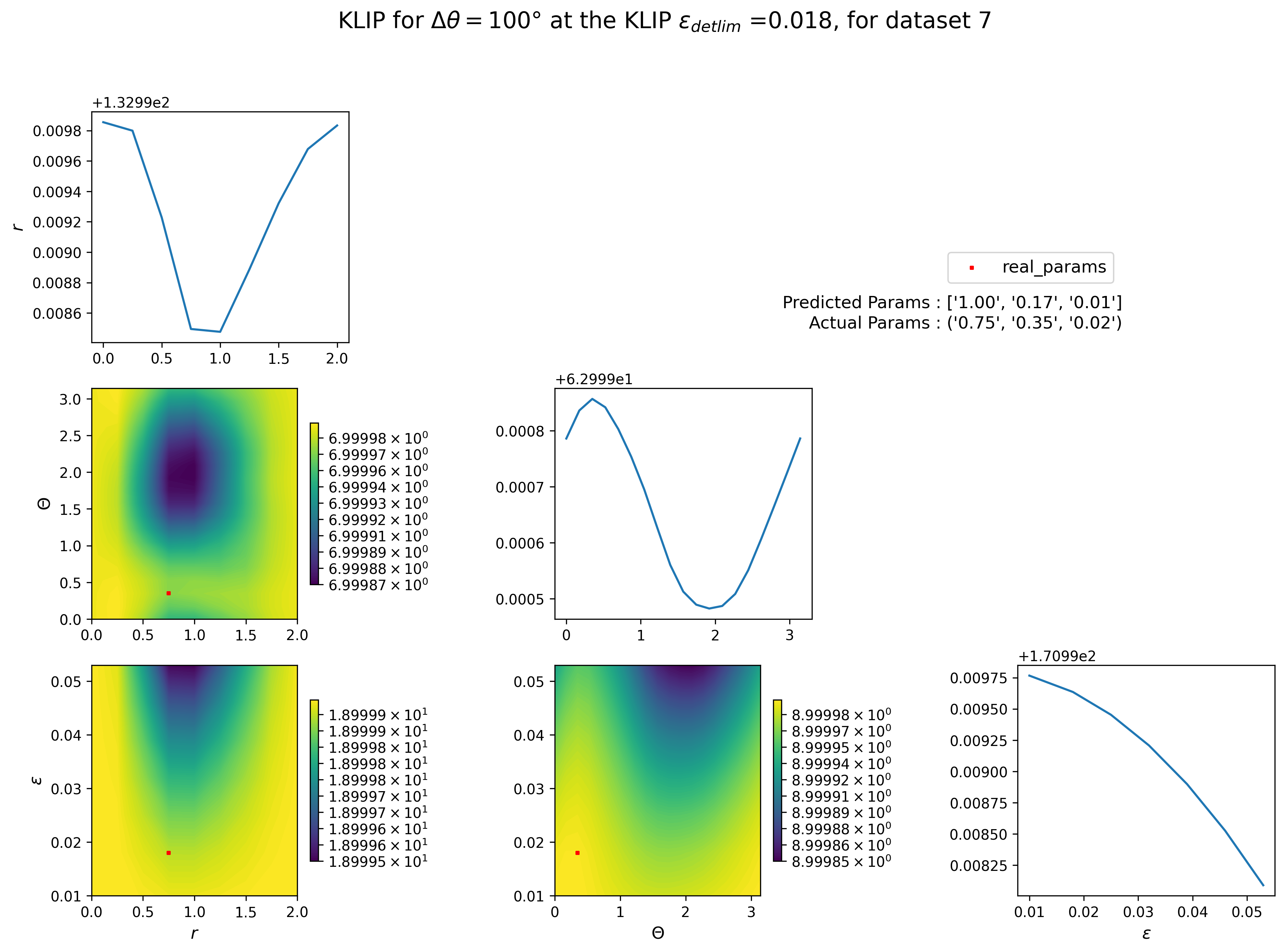}
\caption{Corner plot showing inaccurate localization by standard KLIP subtraction at $\Delta \theta = 100^{\circ}$ and $\epsilon_{KLIP}$. The Antiplanet KLIP method, however, localizes correctly at these parameters as it does for all other datasets at $\Delta \theta = {100^{\circ}, 150^{\circ}}$.}
\label{fig:Appendix KLIP 150 bad}
\end{figure}

Fig.~\ref{fig:100 deg localization example} shows the localization result for the Antiplanet KLIP method summarized in Table~\ref{tab:klip_detlim_localization_25_100} at $\Delta \theta = 100^{\circ}$. Additionally, Fig.~\ref{fig:Appendix KLIP 150} presents the localization at $\Delta \theta = 150^{\circ}$ for the KLIP detection limit ($\epsilon_{KLIP}$). For comparison, while standard KLIP subtraction also localizes the planet accurately for this specific dataset, it fails for other stellar speckle realizations with the same parameters. An example of such a failed localization is shown in Fig.~\ref{fig:Appendix KLIP 150 bad}.

\vspace{2ex}\noindent\textbf{Suvinay Goyal} is an undergraduate student at UIUC, where he works in Experimental Cosmology under the guidance of Prof. Jeffrey Filippini. He completed SURFs at Caltech in 2024 in the Exoplanet Technology Lab under the mentorship of Dr. Yinzi Xin, Prof. Dimitri Mawet, and Dr. Nemanja Jovanovic, and in the Observational Cosmology lab in 2025 under the mentorship of the late Dr. Kenny Lau and Prof. James Bock. He will begin his Physics PhD at Caltech in 2026.  
\end{spacing}
\end{document}